# Dual Fast-Cycling Superconducting Synchrotron at Fermilab and a Possible Path to the Future of High Energy Particle Physics

**Henryk Piekarz**

*Fermi National Accelerator Laboratory*
*Batavia, Illinois 60510, USA*
*E-mail*: hpiekarz@fnal.gov

ABSTRACT: We briefly outline shorter and longer term physics motivation for constructing a dual, fast-cycling superconducting synchrotron accelerator (DSFMR - Dual Super-Ferric Main Ring) in the Tevatron tunnel at Fermilab. We discuss using this accelerator as a high-intensity dual neutrino beam source for the long-baseline neutrino oscillation search experiments, and also as a fast, dual pre-injector accelerator for the VLHC (Very Large Hadron Collider).

KEYWORDS: Accelerators, Synchrotrons, Magnets, Neutrino.

# Contents

## 1. Physics motivation

### 1.1 Brief summary of neutrino mass limit studies

During the past decade developments in the neutrino physics combined with progress in the cosmological models of dark matter and dark energy suggest that neutrinos play a fundamental role in our universe. It has been determined through solar, atmospheric, reactor and accelerator experiments that neutrinos change flavor (oscillate) while passing through matter. This implies that at least two neutrino species have a non-zero mass [1], thus being in a striking contradiction to the Standard Model (SM), and therefore suggesting existence of the physics Beyond the Standard Model (BSM). In addition, the possibility of neutrinos having a small mass may provide a bridge (via e.g. a see-saw mechanism) to the GUT (Grand Unified Theory – physics models where at energies above $10^{14}$ GeV the electromagnetic, weak nuclear and strong nuclear forces are fused into a single field) theories including the origin of mass in the universe. As a consequence of this new situation a need for the resolution to the neutrino physics has risen to a level that is not just complimentary to other high-energy particle physics programs but turns out to be absolutely necessary to further understanding of the microscopic structure and workings of the universe.

In neutrino physics phenomenology neutrinos with physical flavors, $\nu_\alpha$ ($\alpha$ = e, μ, τ), are assumed to be linear super-positions, through a unitarity matrix, of neutrino fields with definitive masses $\nu_i$ (i = 1, 2, 3). A common parameterization for this matrix uses mixing angles, $\theta_{ij}$ = (0, 2π) typically represented by $\sin^2\theta_{ij}$, and a CP-violating phase $\delta_{CP}$ = (0, 2π). The current neutrino phenomenology also implies that two of the neutrino species have relatively close masses while the mass of the third one is either much heavier (normal hierarchy) or much lighter (inverted hierarchy) of the "doublet". The lightest (heaviest) neutrino in the doublet is called $v_1$ ($v_2$) and their squared mass difference is defined as $\delta m^2 = m_2^2 - m_1^2 > 0$. The mass difference between $m_3$ and $m_{1,2}$ doublet is defined as $\Delta m^2 = |m_3^2 - (m_1^2 + m_2^2)/2|$. The recent global analysis [2] of solar, atmospheric, reactor and accelerator neutrino data projects that within a 2σ boundary the $\delta m^2 = (7.92^{+0.09}_{-0.09}) \times 10^{-5}$ eV$^2$, and the $\Delta m^2 = (2.4^{+0.23}_{-0.23}) \times 10^{-3}$ eV$^2$ implying that mass of at least one of the neutrino species is likely to be in the range of (0.01–0.05) eV. We should point out that in another neutrino data analysis [3] this neutrino mass is (0.04 - 0.10) eV, and some individual experiments, e.g. [4], set the upper mass limit at (0.3) eV, significantly higher than those from the global fits. The higher mass value is mostly from terrestrial experiments while the lower one comes from the solar neutrino studies. In the analysis [2] the most likely values of the mixing angle parameters are also given with $\sin^2 2\theta_{13} = 0.036^{+0.0.09}_{-0.036}$ (at 2σ uncertainty level), meaning that it can even be very close to zero.

**1.2 Neutrino flavor-change reactions and the magic baseline**

The neutrino mass can not be directly measured, but as neutrinos pass through the matter they can change the flavor, the process that is described as oscillations. The detection of the oscillations would be a manifestation that neutrinos have mass. The probability of the oscillation is a function of all the mixing angles and other parameters, so the potential smallness of the $\sin^2 2\theta_{13}$ parameter has a strong impact on the probability of the oscillation, and consequently on the feasibility of the experiment to detect neutrino mass. In addition, the complexity of the oscillation function produces typically up to eight-fold degenerate solutions to the experimental data, adversely affecting oscillation detection thresholds. As example of how degeneracy of theory parameters affects sensitivity of the experiment we show in figure 1 the predictions for the recently proposed NOvA experiment [5] at Fermilab. One can see that combination of the $\delta_{CP}$ degeneracy with that of the $\Delta m^2$ widens the projected neutrino oscillation detection thresholds in terms of the $\sin^2 2\theta_{13}$ by more than a factor of 2.

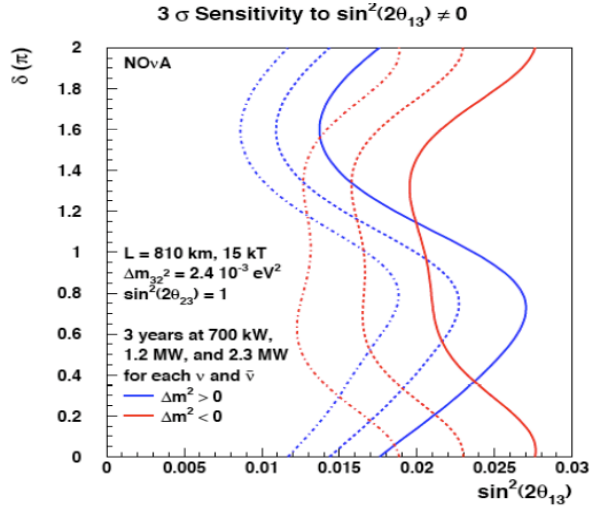

Fig.1 Sensitivity of NOvA experiment to $\sin^2 2\theta_{13}$ as a function of $\delta_{CP}$,
and for both the negative and the positive sign of $\Delta m^2$.

It has been shown recently [6], however, that there is an experimental condition when the degeneracy induced by the theory parameters can be strongly suppressed for the $\nu_e \to \nu_\mu$ or $\nu_\mu \to \nu_e$ appearance probability in matter. This appearance probability, $P_{e\mu}$, can be expanded in the small hierarchy parameter $\alpha = \Delta m_{21}^2 / \Delta m_{31}^2$ and the small parameter $\sin 2\theta_{13}$ as shown below:

$$P_{e,\mu} \sim \sin^2 2\theta_{13} \sin^2\theta_{23} \sin^2[(1-A)\Delta]/(1-A)^2$$
$$+/- \; \alpha \sin 2\theta_{13} \xi \sin(\delta_{CP}) \sin(\Delta) \sin(A\Delta) F(A, A\Delta)$$
$$+ \; \alpha \sin 2\theta_{13} \xi \cos(\delta_{CP}) \cos(\Delta) \sin(A\Delta) F(A, A\Delta)$$
$$+ \; \alpha^2 \cos^2\theta_{23} \sin^2 2\theta_{12} \sin^2(A\Delta) / A^2 \qquad (1)$$

where $\Delta = \Delta m_{31}^2 L / 4E$, $\xi = \cos\theta_{13} \sin 2\theta_{12} \sin 2\theta_{23}$, and $A = +/- (2\sqrt{2} G_F n_e E) / \Delta m_{31}^2$. The L is the baseline for the neutrino oscillation, and E is the neutrino energy. The $G_F$ is the Fermi coupling constant and the $n_e$ is the electron density in matter. The sign of the second term is determined by choosing either $\nu_e \to \nu_\mu$ (positive), or $\nu_\mu \to \nu_e$ (negative) oscillation channel in the formula (1). One can see that for the $\sin(A\Delta) = 0$ all but the first term disappear. This condition is for a nontrivial solution with $\sqrt{2} G_F n_e L = 2\pi$, or in terms of constant matter density, $\rho$, equivalent to a magic baseline, $L_{magic}$:

$$L_{magic} \text{ [km]} = 32726 \; 1/\rho \text{ [g/cm}^3] \qquad (2)$$

With a standard value of $\rho = 4.3$ g/cm$^3$ the magic baseline is $\sim 7630$ km, but with the PREM (Preliminary Reference Earth Model) density it is $\sim 7250$ km long. Naturally, the experiment at magic baseline alone does not allow for measurement of other parameters. So, data from a second baseline are needed to fulfill this void. The authors of reference [6] provide analysis strongly suggesting that a combination of data from the magic baseline with those from the one of $\sim 3000$ km length would allow for the determination of the neutrino mass hierarchy (sign of $\Delta m_{13}^2$) and the CP violation ($\delta_{CP}$ phase) down to the values of the $\sin^2 2\theta_{13}$ parameter by several orders of magnitude below of those of projected in any current experiment.

**1.3 Fermilab site as point of origin for the neutrino experiments with magic baselines**

Following the suggestion of the magic baseline of 7250 km length a proposal for sending a neutrino beam from CERN to India has been put forward [7]. For the Fermilab site a neutrino beam has to be sent to the central Europe (e.g. FNAL->Gran Sasso, IT) for the magic baseline.

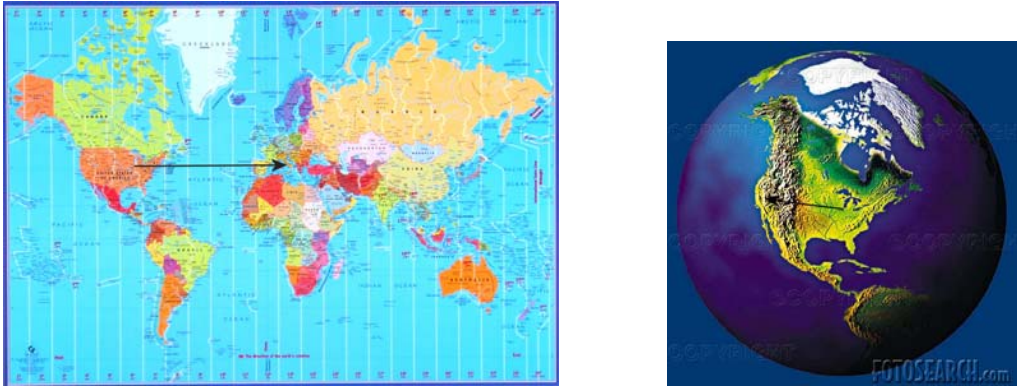

Fig.2 Neutrino beam paths: FNGS to Gran Sasso (left) and FNMW to Mt Whitney (right)

Fermilab site is also well located for the matching baseline of ~ 3000 km length if a neutrino beam is send to the western part of the US (e.g. FNAL -> Mt Whitney, CA). The geographical paths for the FNGS and FNMW baselines are shown in figure 2. The Gran Sasso is naturally of a great interest because the CNGS detector is already residing there. The Mt Whitney is also an interesting location because it is a tall (4300 m) mountain of a granite rock in a non-seismic area. A cave near the ground level would be sufficient for housing the detector there. The other potential sites at about 3000 km distance from the Fermilab are: San Jacinto, CA and Icicle Creek, WA. Both these sites are being considered at present for the National Underground Laboratory. We note that all the above locations are in the vicinity of the west-coast US universities and HEP institutions that would be certainly interested in designing, building and operating the neutrino detectors there.

The path of neutrinos through the Earth's crust for the FNGS and FNMW experiments is shown in figure 3. The maximum depth of the neutrino beam into the Earth's crust is ~1660 km for the FNGS and ~185 km for the FNMW. For comparison, the maximum depth of the neutrino path to the MINOS, Nova and CNGS experiments is about 10 km (all three experiments have baselines of ~ 735 km length). The much greater averaged depth of the neutrino path for FNGS and FNMW experiments has the advantage of a more uniform and better predictable Earth's matter density which in turn will help to project the neutrino interactions while they are passing through the Earth's crest.

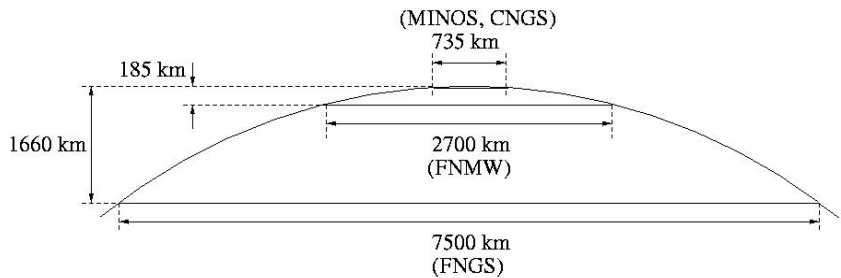

Fig.3  FNGS and FNMW neutrino paths through Earth's crest as compared to MINOS/CNGS

**1.4 Possible new particle mass scale and reach of various colliders in post-LHC era**

The determination of existence (or non-existence) of the Standard Model Higgs is the most important high-energy particle physics goal at present. The LHC is very well set to discover and investigate Higgs up to mass of 0.8 TeV, which is nearly an order of magnitude more than the 90 GeV mass of the Standard Model highest likelihood. Already the results from the Tevatron suggest that the lower limit of the Higgs mass is likely to be above 150 GeV, and therefore on the fringes of acceptability within the Standard Model.  As indicated in Chapter 1.1 the past decade developments in the neutrino physics combined with the cosmological theories suggest that neutrinos have mass, a hypothesis that can not be accommodated within the Standard Model. Although the Standard Model Extensions (e.g. Super-Symmetry) are vigorously being developed, a stronger and stronger consensus is being built that one should expect physics Beyond the Standard Model (BSM) which may better explain the workings of the universe.

Naturally one would like to know the energy scale at which this new physics should occur. It is believed that neutrino physics sheds important light on this matter. As the neutrinos do not carry charge (unlike other fermions) it is possible to assume that at the origin of the Universe they were created as a particle doublet with masses split by many orders of magnitude. As we know that one partner of this doublet has a very low mass the other one would have to have a very large mass to satisfy relation (3) between the doublet particle masses and the weak force, v, that enabled their creation:

$$N \times m_v = \text{const.} \times v^2 \qquad (3)$$

The $m_v$ is a neutrino mass and the heavy neutrino, N, is presumed to be right handed Majorana particle. Based on this assumption a model-independent upper bound on the scale, $\Lambda_{Maj}$, of the Majorana-neutrino mass generation was projected [8] as shown in formula (4):

$$\Lambda_{Maj} = 4\pi v^2 / \sqrt{3}\, m_v \qquad (4)$$

where $v = (\sqrt{2}\, G_F)^{-1/2} \sim 246$ GeV is the SM weak scale based on the Fermi coupling constant, $G_F$. The formula (4) was derived using the presumed neutrino scatterings to W and Z gauge bosons and thus being specific to the Standard Model. Using this formula we plot in figure 4 a relation between neutrino masses, $m_v$ and N, bound to the observed limits of the mass $m_v$. The current $v_e$ mass limits are the lowest ones and therefore they set the highest possible N mass range of $(10^{16} - 10^{17})$ GeV. The LHC is expected to search for the Higgs of up to 0.8 TeV mass relating possibly to a weak force of 1.6 TeV. As the mass N of the Majorana neutrino increases with the square of the weak force its mass reach would then extend up to 40 times $((1.6/0.25)^2 \approx 40)$, and therefore get into a very close vicinity of the Planck Scale. The constant in equation (3) can not be predicted neither for the Standard Model Extensions (as these particles have not yet been observed), nor for the Beyond Standard Model theories which has not been even defined yet. Nevertheless, it seems plausible that regardless of the particle model the weak scale does not need to be much higher than ~ few TeV in order for the Majorana neutrinos reach the ultimate mass range of $(10^{18}-10^{19})$ GeV. Indeed, many theorists [9] believe that the new physics will open at a mass range of $\geq 1$ TeV. This is an interesting observation because it suggests that the mass reach of the future colliders does not need to be much higher than that of the LHC to investigate physics of the Standard Model Extensions, or of the Beyond Standard Model theories. This is a primary reason for considering the DLHC (Double Energy of LHC) for the LHC upgrade. Building, however, an entirely new accelerator with mass reach of (5-10) TeV (rather than (1-2) TeV) is a much safer option. The new particle mass reach in terms of the achievable maximum energy per parton at existing, and considered for the future, colliders is shown in figure 5. As the DLHC accelerator will be placed in the LHC tunnel its energy increase can only be achieved by doubling magnetic field of the LHC magnets, a magnet technology yet to be developed. For the VLHC (Chapter 3.1) we propose a circumference 4 times longer than that of the LHC thus allowing for 4 times higher energy (VLHC I) using the existing (e.g. LHC-type) magnet technology, or 8 times higher (VLHC II) with a magnet technology anticipated for the DLHC.

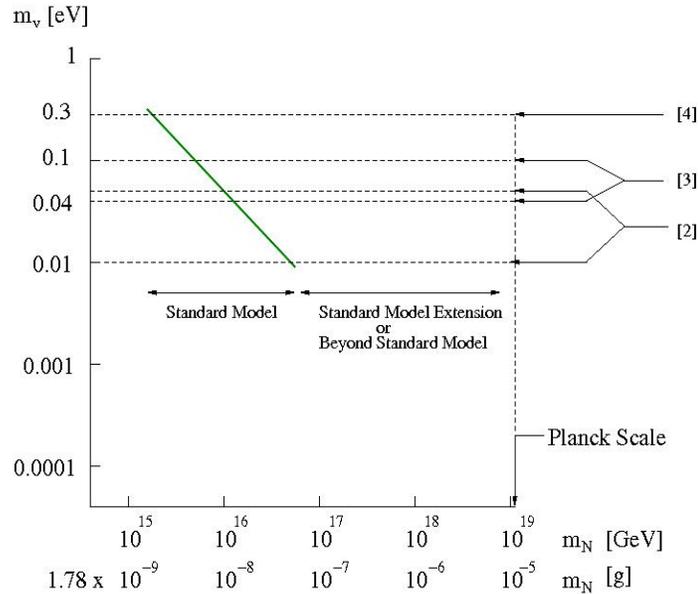

Fig. 4 Majorana neutrino mass N versus neutrino mass $m_v$ limits as given in Refs [2,3,4]. The equation (4) with a Standard Model weak scale of 0.25 TeV was used.

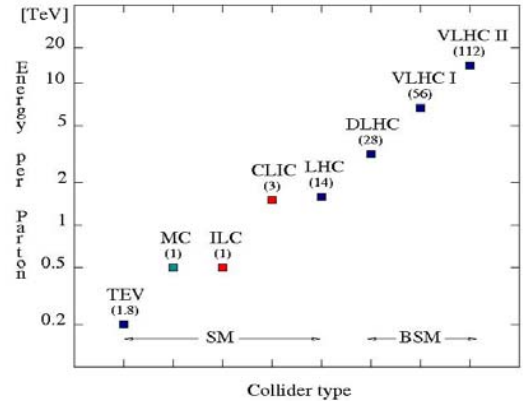

Fig. 5 Energy per parton at various colliders (cms energy is in bracket):
TeV – pbarp, MC- $\mu\mu$, ILC, CLIC – ee, LHC, DLHC, VLHC – pp.

## 2. DSFMR as high intensity neutrino source

### 2.1 Overview of the proposed accelerator complex with DSFMR

We expect high intensity neutrino source at Fermilab to fulfill the following conditions:

- Allow for non-interrupted operations of MINOS and NOvA experiments.
- Produce simultaneously two neutrino beams of intensity to be satisfactory for 7250 km and 3000 km baseline experiments.
- Construct long baseline neutrino production lines fully within the Fermilab property.
- Maximize use of the existing Fermilab accelerator infrastructure to minimize the cost.
- New accelerator complex and its neutrino beam lines should be completed in a period of time that the followed-up physics program will be viewed as much advanced relative to the similar programs elsewhere.

The Dual Super-Ferric Main Ring (DFSMR) accelerator as proposed in [10] fulfills well the above conditions. The outline of the proposed new Fermilab accelerator complex with DSFMR is shown in figure 6. When the Tevatron stops its operations it would be replaced with two rings of fast-cycling synchrotrons (DSFMR) placed in the Main Ring tunnel.

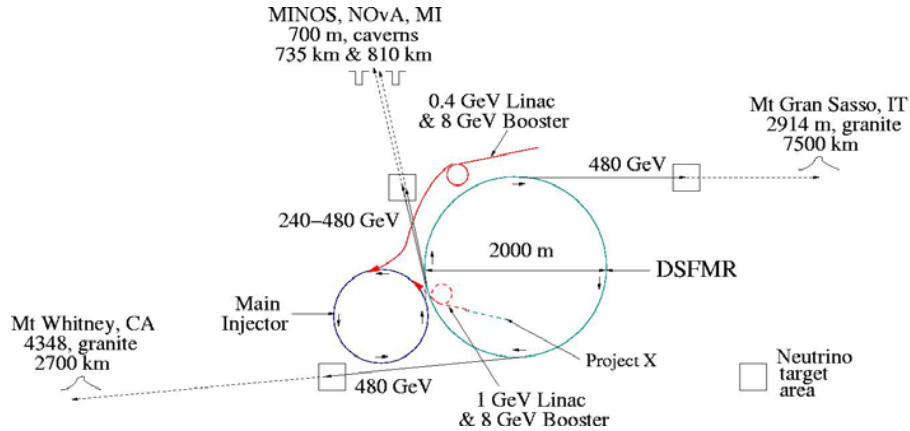

Fig. 6 Proposed arrangement of DSFMR at the Fermilab accelerator complex

As the Main Injector is not only the latest addition to the Fermilab accelerator complex but it is also a well functioning fast-cycling synchrotron it constitutes firm part of the proposed new Fermilab accelerator complex. The main three components of the new accelerator complex are then: Pre-injector (8 GeV), Main Injector (8-120 GeV), and in the final stage the DSFMR (120-480 GeV). The present Pre-injector consists of 0.4 GeV Linac and 8 GeV Booster that can be used as is for the start up. If the Project X [11] is implemented this Pre-injector would become the 8 GeV H$^-$ Linac with a Booster serving only as the electron stripper ring.

The important part of the DSFMR proposal is that this machine would consist of two accelerators rings embedded in the Main Ring tunnel. Using two accelerators instead of one allows operate two neutrino beams simultaneously, a much preferred option for the two (2700 km and 7250 km) long baseline experiments. Ability to cross-check data between these very difficult experiments would greatly benefit in debugging problems associated e.g. with a very low rate of events and consequently improve physics analysis. In addition, until the new long baseline experiments become operational MINOS and NOvA experiments could not only benefit from much higher beam intensity but also operate simultaneously.

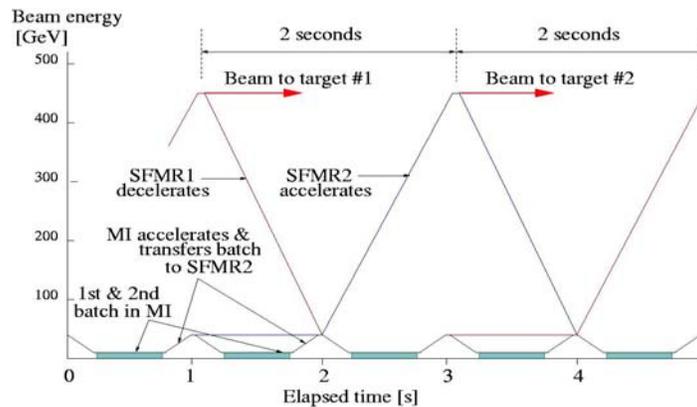

Fig. 7 Time sequence for beam stacking, ramping and extraction onto neutrino production targets with DSFMR

The time sequence for beam stacking in the Booster, MI and DSFMR accelerators together with their respected ramping times and beam extractions is shown in figure 5. The first set of proton pulses from the Linac is stacked in the Booster and then accelerated to 8 GeV. This beam batch is then transferred to the Main Injector, accelerated immediately to 48 GeV energy and then transferred to one of the DSFMR rings where it will await for a second proton beam batch from the Main Injector to arrive. The DSFMR ring will accelerate both batches up to 480 GeV and then extract them into one or two neutrino beam production lines, as desired. The timing sequence with the 8 GeV H$^-$ Linac of the Project X as the Pre-injector remains the same except that the proton beam intensity is achieved through Linac cycling rate combined with a much higher current of the H$^-$ ion source.

By using the Main Ring tunnel for DSFMR the existing Tevatron infrastructure, e.g. wall power distribution, RF system, cryogenic distribution lines, etc. will be re-used. The RF system will have to be modified and expanded but importantly it will be shared by both accelerators which are timed to operate interchangeably. The Main Ring tunnel currently houses two accelerator rings (Main Ring and Tevatron) with the Main Ring magnet alone being larger than the dual magnet assembly proposed for the DSFMR (see Chapter 2.2), so the space in the Main Ring tunnel is not an issue. Some of the infrastructure will have to be modified and expanded, e.g. cryogenic support system for two accelerator rings instead of one. We are aiming in our DSFMR magnet design that the two accelerator rings will use no more cryogenic power than at present used by the Tevatron, so no construction (and operation) of a new cryogenic plant will be needed. A very tentative cost estimate of DSFMR presented in [12] suggests that it should be possible to construct two DSFMR accelerator rings in Main Ring tunnel at the cost similar to that of the 8 GeV superconducting linac of the Project X [11].

## 2.2 Fast-cycling superconducting transmission line accelerator magnets

The use of fast-cycling superconducting magnets for the DSFMR accelerator is the most important element of this proposal. The normal-conducting fast cycling magnets have been successfully used in large synchrotrons (e.g. 120 GeV Main Injector at Fermilab). In order to achieve the required B-field in the 2 Tesla range these magnets have to operate in the super-ferric regime with a core determining the strength of magnetic field in the gap. Consequently, both the size of the conductor and the desired width of the magnet gap determine the overall size of the core. The use of the superconducting magnets allows minimization of the magnet cross-section area by nearly an order of magnitude due to only a small space required for the superconductor winding which can typically carry current of ~ 40 times higher density than the copper conductor.

The concept of transmission line superconducting accelerator magnets was first developed for the VLHC Stage 1 accelerator [13], and a short prototype of the VLHC-LF magnet was successfully built and tested [14,15]. This magnet has been also proposed recently for the Low Energy Ring [16] at the LHC. We are expanding on the VLHC-LF magnet experience to design a fast-cycling transmission line superconducting accelerator magnet operating in the regime of 2 T/s with the repetition rate of 0.5 Hz, as it is required for the DFSMR. The use of a transmission line conductor to power the magnet facilitates application of the combined function magnet design simplifying the overall accelerator design, but as shown in VLHC-LF [13] a set of corrector magnets placed periodically around the ring is needed. We anticipate these correctors to be superconducting magnets as well.

The main issue for the DSFMR accelerator is reduction of the magnet superconductor power losses to the level that the required cryogenic power will be cost-wise acceptable. In our magnet design the superconductor operates at the liquid helium temperature while the magnetic core is cooled to a liquid nitrogen temperature. The cost of the heat loss recovery for the liquid helium is about 15 times higher than that for the liquid nitrogen. Consequently, the helium cryogenic power losses in the accelerator magnet string are of the main concern. In past decade there was a considerable effort to build the fast-cycling magnets based on the NbTi superconductor strands. The best to-date AC power losses measured [17] in such magnets are ~ 28 W/m at 4 T/s at maximum B-field of 2 Tesla. This result approximately scales to ~ 7 W/m at 2 T/s with 0.5 Hz cycle. For the two DSFMR rings of ~7000 m each this power loss translates to ~ 96 kW of required helium cryogenic power, being then 4 times larger than the Fermilab cryogenic plant of 24 kW. This is unlikely to be acceptable for the DSFMR project not only due a very high cost of new high-power plant construction but even more so due to high operational cost over the lifetime of the accelerator. Consequently, if the DSFMR project is to succeed a solution has to be found to the fast-cycling superconducting magnets allowing reduce the liquid helium power consumption at least by a factor of 4 to 5. For this reason we are exploring possibility of using the HTS type superconductor to construct the fast-cycling magnet power cable.

In fast-cycling superconducting magnets hysteretic and Eddy current losses dominate cable power losses. In a superconductor of width $w$ exposed to a perpendicular magnetic field of amplitude $H_m$, hysteretic loss per unit volume of superconductor $Q_h$ per cycle is approximately expressed by equation (5) [18]:

$$Q_h \approx \mu_0 \, w \, J_c \, H_m \qquad (5)$$

where $\mu_0$ is the permeability of free space and $J_c$ is the superconductor critical current density at the $H_m$ field. For a given magnetic field and the critical current the hysteretic loss can only be reduced by minimizing the filament area exposed to the magnetic field. The HTS (High Temperature Superconductor) can be manufactured in a form of tape containing only a single superconductor filament. Typically this filament is wide (4 mm) but very thin ($\leq$ 2 μm) with the HTS tape carrying superconducting current of a similar strength to that of the LTS (Low Temperature Superconductor) wire strand of the same overall cross-section. In the magnetic design the tape shape of HTS superconductor can be exploited to minimize the superconductor area exposed to the magnetic field by positioning the narrow edge of the tape toward magnetic field. The tapes can be stacked together (figure 8) with wide surfaces facing each other and with the conductor stack oriented in a way that the wide surfaces of the filaments are parallel to the magnetic field. This arrangement not only minimizes conductor area exposed to the magnetic field but the parallel orientation of the magnetic field to the wide surface also maximizes the filament's critical current.

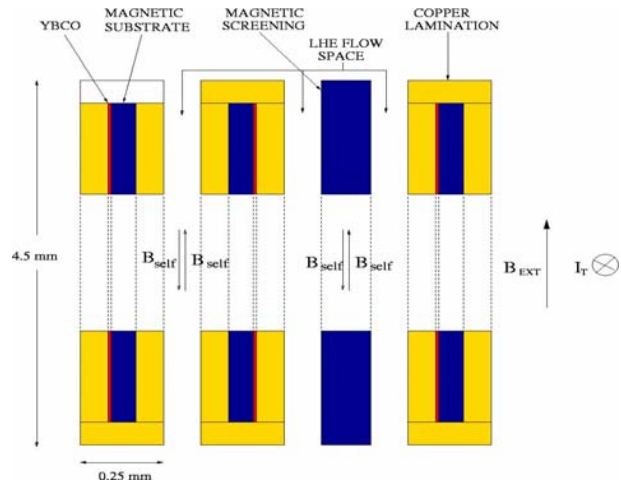

Fig. 8 Arrangement of 344S HTS tapes in a conductor

The second most important power loss for the superconductor is caused by the coupling between the transport current induced magnetic self-fields in the neighboring filaments. This power loss if not suppressed may saturate to the level of the hysteretic one. As part of the fabrication process the HTS tape is mounted on a magnetic substrate. This substrate may increase the Eddy current losses but it was shown in [19] that by arranging the neighboring tapes with magnetic substrates facing each other the AC transport current losses are suppressed to a minimal level known as the Norris elliptic curve. This effect is attributed to canceling of self-fields in the substrates. Following this result we also add an additional magnetic substrate between the HTS tape pairs, and in this way further minimize the AC transport current induced losses.

The arrangement of the conductor stack inside the cryogenic pipe is shown in figure 9. The stacks of the HTS tape pairs are assembled with a narrow space between them allowing for helium coolant to flash wide surface of the tapes and providing in this way a very efficient heat absorption. The space between tape pairs is made with a help of staggered Kapton tape rings. This in turn provides electrical isolation between the pairs helping to suppress further the transport current induced coupling. The split cryogenic pipe which holds conductor stack will be welded under ~ 1 ton/m tension to prevent the stack movements when the conductors are energized and the sweeping magnetic field is present. The magnet conductor will be constructed of multiple stacks summarily providing the required transport current $J_t$ for the designed B-field in the beam gap of the magnet. Magnetic core of the window frame shape as shown in figure 10 is very suitable for the HTS conductor application. The magnet conductor stack will be arranged in a common cryostat.

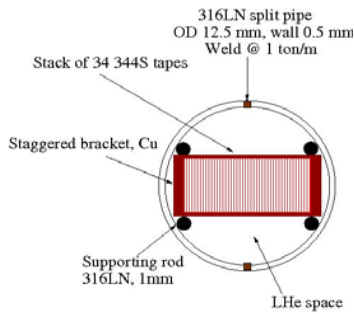

Fig. 9 Arrangement of the test HTS stack in cryogenic pipe

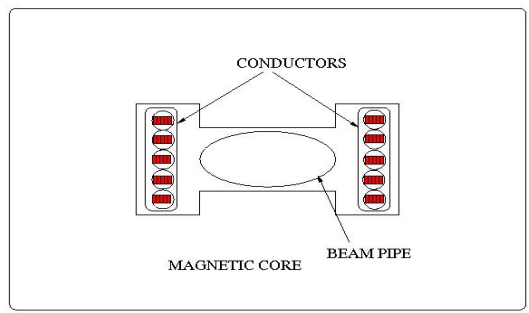

Fig. 10 Schematic view of a window frame magnetic core with the HTS conductors

In order to minimize occurrence of the quench the ratio of the $J_t/J_c$ should be as high as practically possible but due to conductor cost typically $J_t/J_c$ is ~ 2. The selection of the $J_t/J_c$ ratio depends also strongly on the allowable helium coolant temperature margin to conduct the quench-free operations. For the 344S conductor used for this study the $J_c$ falls only by a factor of two [20] between 5 and 25 K at the 2 Tesla field leading then to a 20 K safe operation margin if the number of used strands is based on the $J_c$ at 25 K.

In a magnetic core of a window frame shape the magnetic flux orientation in the gap is designed to produce a dipole B-field, $B_y = B_{max}$ with $B_x \sim 0$. In practice the $B_x$ value is typically ~ 0.1% of the $B_{max}$ and the B-field in the conductor space can deviate even more from the parallel orientation of magnetic flux lines. As mention earlier the sweeping B-field power induced losses in the HTS tape can be reduced if the magnetic field is parallel to the wide surface so only tape's narrow edge is exposed to the magnetic field. This means that magnetic core design should aim at $B_x$ values to be as small as possible across the conductor stack. A very preliminary magnetic design [21] of a DSFMR dipole with 40 mm x 80 mm gap and 2 Tesla field is shown in figure 11. The average $B_x$ component of the magnetic flux crossing the HTS conductor stack area (as indicated in figure 11) was calculated to be ~ 350 G. In order to estimate fully the HTS strand power losses the detailed structure of the tape must be considered. Such an analysis was done for the YBCO type superconductor in [18] and then expanded in [22] to the 344S-2G (YBCO) tapes proposed for the DSFMR magnets. In addition to the YBCO hysteretic losses there are Eddy current losses in the 344S-2G tape lamination and the magnetic substrate. As the magnetic substrate saturates at ~ 0.25 Tesla its Eddy losses primarily scale with the cycling frequency and not the B-field itself. Using the analysis presented in [18, 22] we project YBCO hysteretic losses with the $B_x = 350$ G, $B_y = 2$ Tesla and operation cycle of 0.5 Hz at ~ 1.1 mW/m, and with the contribution of the eddy currents in the magnetic substrate and the copper lamination the total projected power loss is ~ 1.3 mW/m. For the magnetic core design the B-H response of the Si3%Fe 100 μm thick laminations as supplied by the Mapes & Sprowl Steel was used. With this steel the 2 Tesla field in the 40 mm gap is achieved using a transport current of ~ 62 kA, and so 70 kA was assumed for the practical transport current in application for the DSFMR magnet. This transport current based on the data in [20] requires 204 of the 344S-2G tapes per magnet for $I_t = 0.5\ I_c$ @ 5K ($I_t = I_c$ @ 25K) and with 2 Tesla B-field crossing the conductor. Consequently, the total projected dynamic power loss in the magnet conductor for 2 T/s at 0.5 Hz cycle is ~ 265 mW/m. The estimated static losses with a transmission line type conductor are ~ 360 mW/m of magnet [22]. So, very tentatively the total power loss is ~ 625 mW/m, or ~ 9 kW for the DSFMR accelerator with the dual rings of 7 km circumference each. This is much less than 24 kW of the available Tevatron cryogenic power, giving considerable margin to counter uncertainties in our projections.

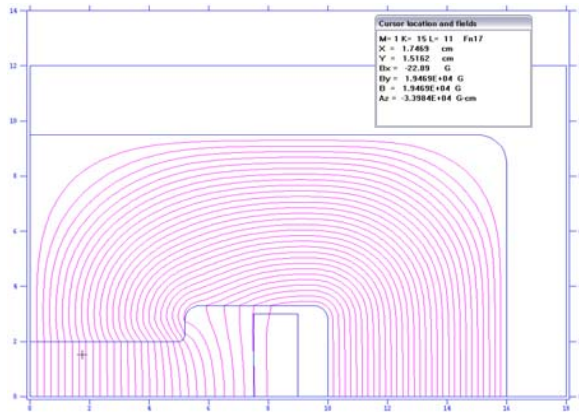

Fig. 11 Lamination design of DSFMR dipole. Box in the gap indicates conductor stack area.

A disparity of the power loss between the measured NbTi strands based cable and the projected above using the 344S HTS tapes may be in part due to a very fundamental difference the way the two type of cables are made. The NbTi strands used in [17] are based on the 0.5 mm diameter wire (the smallest technically available) filled with 1000's of micron size filaments embedded in the copper matrix. The small size of the filaments helps to minimize the hysteretic losses but the copper matrix gives rise to significant Eddy current induced losses. The filaments are arranged into the twisted pairs which helps to minimize the AC current induced self-field coupling but the lack of electrical isolation impedes this effort. This matrix is needed to hold filaments together into a wire-like strand but it also allows sharing the superconducting current to stabilize the strand against the heating disturbance. The operational maximum temperature margin for the NbTi strands is typically much less than 2 K, so having the liquid helium to the strand contact area as large as possible is of great importance. But this is difficult to achieve with the wire strands as they must be held under a high tension to prevent any movement caused by the force of the sweeping magnetic field. Such movement will cause heating of the strand thus making the conductor easily prone to quenching within its small operational temperature margin. As a result the cooling helium contact area in the NbTi cables is rather small. Contrary to the NbTi cable, in the proposed above 344S HTS one, there are no strands but only the individual filaments which pairs are electrically isolated from each other, and a 4 mm wide surface on both sides of the tape pair is being flashed with the cooling helium.

In summary, possible application of the 344S-2G conductors to energize fast-cycling magnets is promising but this concept must be tested for unaccounted adverse effects that may considerably increase required liquid helium cooling power. For this purpose a test is being assembled in the E4R enclosure at Fermilab. A sketch of this test arrangement is shown in figure 12, and a preliminary arrangement of the test conductor assembly is shown in figure 13.

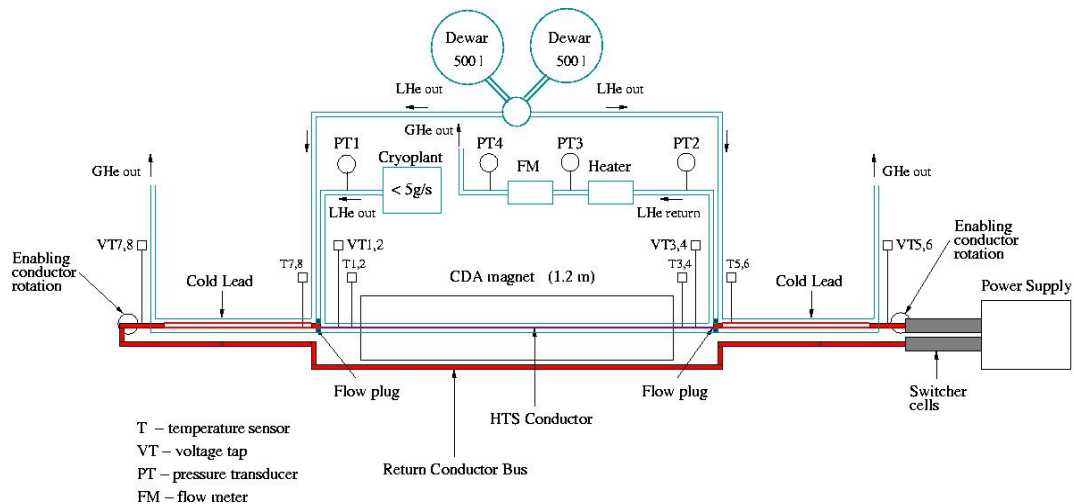

Fig. 12 Sketch of the HTS conductor test arrangement

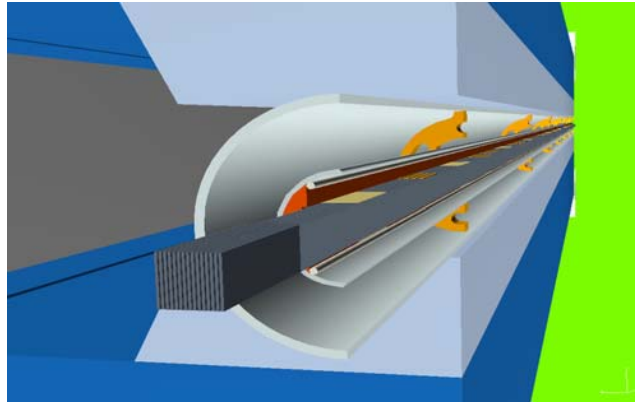

Fig. 13 A cross-section of the test HTS conductor stack inside a dipole magnet gap: black- HTS tapes, red- conductor stack holder, orange- cold pipe support and grey- inner cold and outer cryostat pipes

The goal of the test is to measure power losses in the conductor sample consisting of 34 HTS tapes exposed to an external sweeping magnetic field of up to 3 T/s. The dipole magnet of $B_{max}$ = 0.7 Tesla and repetition rate up to 3 Hz will be used in these tests. The test conductor will be rotated relative to the direction of the field lines in the magnet gap thus exposing the wide surface of the HTS tapes to a varying strength of the $B_x$ component of the magnetic field. In addition, the test conductor will be also energized with the external current source to measure power losses induced by a pulsing current. Precision measurements of the temperature gradients as well as the liquid helium pressure drop along the HTS conductor length combined with the measurements of the liquid helium flow rate will allow determine the HTS conductor cryogenic power loss. In the test the conventional power leads will connect the room temperature power supply with the test conductor.

Possible arrangement of two magnetic cores for DSFMR accelerator is shown in figure 14. The cores are placed inside the pipes for support. These pipes serve also as cryostats for cooling the conductor with LHe and the cores with LN2. The B-H response of magnetic core improves with lowering temperature and the static heat loss at the conductor cable minimized. As the cross-section of the DSFMR magnet core is small (32 cm x 20 cm) placing the magnet inside a pipe is a good solution for handle the cores transportation and suppressing deformation.

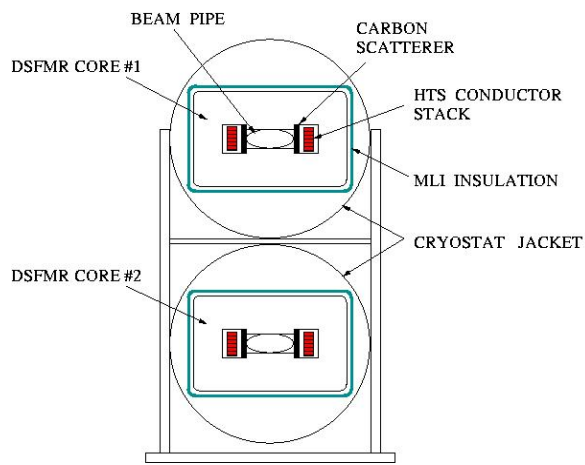

Fig.14 Arrangement of two magnetic cores for DSFMR accelerator

The way the conductor is arranged to energize magnetic core plays important role in determining the effectiveness of the used cryogenic power. The conductor in a magnet can be arranged in two ways: (1) single winding or (2) multiple winding as shown in figure 15. In the multiple winding the conductor cooling channel can be of rather considerable length causing significant helium pressure drop that in turn leads to a non-uniform cooling along the winding. In the transmission line conductor the helium path can be reduced to a single magnet length (or a selected magnet string) as helium supply and return channels run parallel to the magnet. As discussed in [23] such an arrangement provides not only the most efficient cooling of the magnet conductor but by returning the only minimally warmed-up helium to the cryogenic plant it minimizes the overall required cryogenic power.

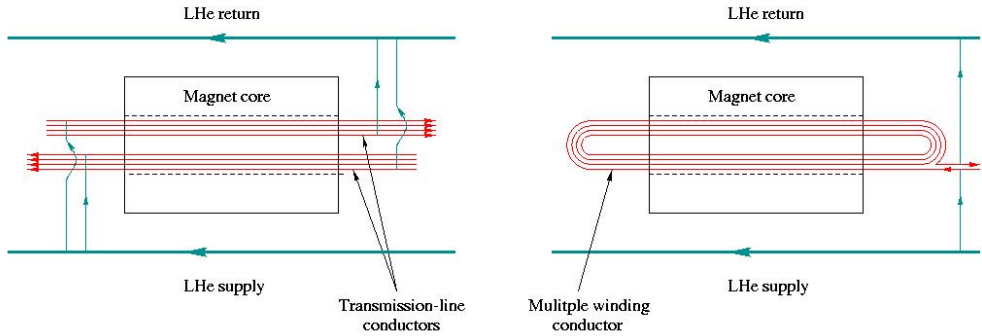

Fig. 15 Single (left) and multiple (right) magnet core winding with a superconductor

## 2.3 Arrangement of long baseline neutrino production lines with DSFMR

The most important feature for the neutrino production lines based on the DSFMR accelerator is that these lines fit very well within the Fermilab proper as indicated in figure 16. A sketch of the neutrino production lines inclining into the earth is shown in figure 17. For the assumed 1000 m long meson decay pipes the required maximum depth is ~240 m for FNMW and ~700 m for FNGS. The excavation and construction of the tunnels constitutes rather great engineering challenge. The fact, however, that these decay pipes are only about 1.5 m in diameter and with no need for a human access along the entire pipe length should help the construction effort and keep the cost at some reasonable level. There will be shafts to the neutrino production target caves, and to the caves at the deep ends of each decay pipe where the identifying neutrino production detectors will be located.

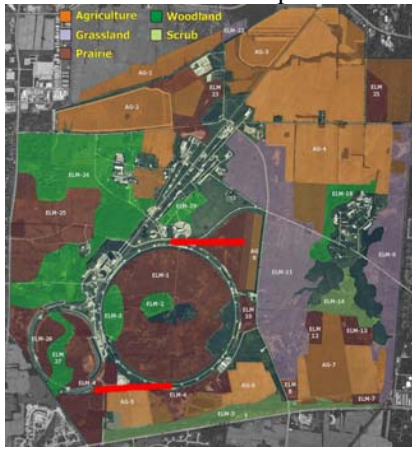
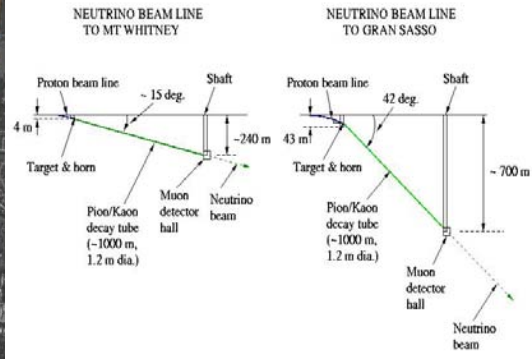

Fig.16 Bird's view of Fermilab proper (neutrino production lines in red)    Fig.17 Vertical inclination of FNMW and FNGS neutrino production lines

**2.4 Estimated sensitivity limits for neutrino oscillation search experiments with DSFMR**

At present the Main Injector allows for proton bunch intensities of $N_b \sim 10^{11}$ without adversely affecting circulating beam phase space due to e.g. electron cloud effects [24]. With the $N_b \sim 10^{11}$ protons per bunch the maximum allowable number of stored protons in the Main Injector is $\sim 5.4 \times 10^{13}$. As the DSFMR ring circumference is double in size of the Main Injector and the beam pipe cross-sections are about the same, one should expect to store $\sim 1.08 \times 10^{14}$ protons in each of the DSFMR rings. The neutrino beam flux is typically measured by beam power on production target which is expressed in the formula (6), where $N_p$ is a number of protons on target in units of $10^{20}$, $E_p$ is the proton energy in units of GeV, and T is the time of exposure in units of $10^7$ seconds.

$$\text{Beam Power [MW]} = (N_p \times 1.62 E_p) / (1000 \times T) \quad (6)$$

For proton beam energy of 480 GeV, cycle time of 2 seconds and with $1.08 \times 10^{14}$ protons per cycle the projected DSFMR beam power on target (POT) is:

$$\text{POT} = (10^{-7} \times 1.62 \times 480) / (1000 \times 2 \times 10^{-7}) = 8.6 \text{ MW} \quad (7)$$

The 8.6 MW exceeds by a factor of 2 the probably acceptable beam power on the neutrino production target (even 4 MW target [25] needs extensive R&D). There is a two-fold solution to this problem: (1) – reduce the beam energy to 240 GeV while keeping the same cycle time, and (2) – split and extract two beam batches from the DSFMR, each batch onto its own neutrino production target. The first option is suitable for operations with only one neutrino experiment, while the second option is suitable for simultaneous operations of two independent neutrino experiments which is in fact a primary reason for the DSFMR proposal. Simultaneous extraction onto two production targets with accelerator cycle time of 2 seconds is equivalent to extracting a beam batch onto one production target every 1 second. The fact that 4 MW beam power can be simultaneously available for two neutrino production targets provides a factor 20 advantage over the current neutrino beam production at Fermilab. With the MI beam intensity acceptable at present ($5-6 \times 10^{13}$ per cycle), the HINS would produce maximum beam power of only 0.8 MW [11]. It is expected that the Project X may provide up to 2.3 MW power but for a single target only. The projected DSFMR beam power also exceeds by factor 2 the future J-PARC and CERN (SPL) upgrades [26].

As the purpose of this note is to provide only a qualitative analysis of what can be achieved with the DSFMR as neutrino beam source we use the neutrino flux for the CNGS experiment with 400 GeV proton beam to project the neutrino flux with the DSFMR. As shown in [27] the projected neutrino flux at the CNGS detector site (735 km from source) is $\sim 7.5 \times 10^{-9}$ $\nu_\mu$ / pot_m$^2$. This makes $\sim 4 \times 10^{-3}$ $\nu_\mu$/p.o.t._m$^2$ at $\sim 1000$ m from the production target (excluding detector acceptance). For the DSFMR this rate increases by the ratio of 480/400 to $\sim 4.8 \times 10^{-3}$ $\nu_\mu$/p.o.t._m$^2$. Assuming $5 \times 10^{13}$ p/s, and $2 \times 10^7$ seconds/y one obtains $\sim 4.8 \times 10^{18}$ $\nu_\mu$/y at $\sim 1000$ m from the production target for each neutrino beam to the far detectors. The 1000 m distance is a typical decay path for $\pi \rightarrow \mu + \nu$ in direct production of a neutrino beam with proton synchrotrons and it can be compared to ~700 m path (one leg of a triangle) assumed for a neutrino production from $\mu \rightarrow e + \nu + \bar{\nu}$ decays in the Neutrino Factory. The π and μ decay paths for the neutrino beam production are illustrated in figure 18 which shows that with full acceptance of π and μ the neutrino beam rates per power on target would have to be the same in both cases. For DSFMR (5 x $10^{13}$ p/s at 480 GeV) the power on target is actually 20% higher than for a Neutrino Factory ($10^{16}$ p/s at 2 GeV). It also appears that the μ-decay pipe is typically assumed to be of about the same cross-section (~ 1 $m^2$) as the π-decay pipe suggesting expectation of a similar emittance growth of neutrino beams for both cases.

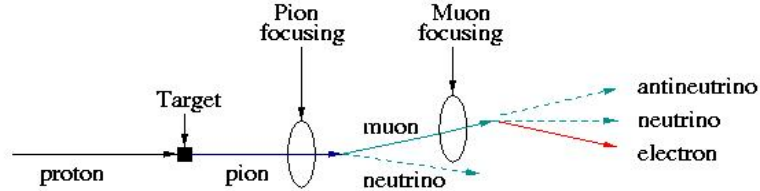

Fig. 18 Neutrino production in one step process (π-focusing, DSFMR), and in two-step process (π-focusing followed by μ-focusing, Neutrino Factory)

Neutrino Factories [28, 29] project typically a useful flux of $\nu_e$ and $\nu_\mu$ neutrinos ~ $10^{20}$ /y with the expectations of ~1.8 x $10^{20}$ $\nu_{\mu,e}$/y (this latter value was assumed for the sensitivity limits estimation in [6]). The most recent plans [30] suggest even a possibility of $10^{21}$ muon decays per year. So, the DSFMR would have the $\nu_\mu$ flux about 400 times lower than that of the ultimate Neutrino Factory. A comparison of projected sensitivity limits with the DSFMR to other accelerator options are shown in figure 19.

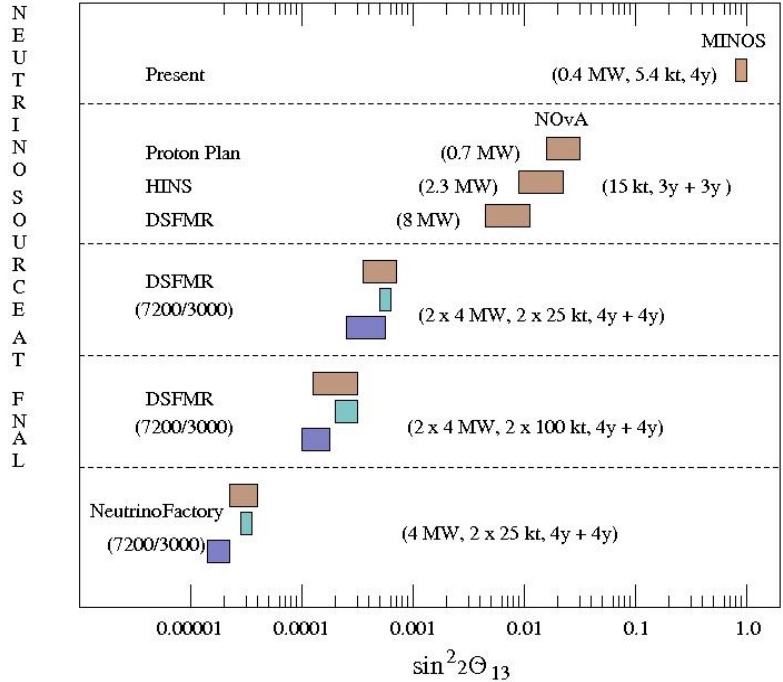

Fig. 19 Sensitivity reaches as function of $\sin^2 2\theta_{13}$ for $\sin^2 2\theta_{13}$ (brown), sign of $\Delta m^2_{31}$ (green), and CP violation (blue) with DSFMR and Neutrino Factory [6,30].

– 16 –

Sensitivity limits scale with the luminosity, L, as $1/\sqrt{L}$. The luminosity is a product of a total neutrino flux and detector acceptance. In order to get a rough approximation of what can be expected with DSFMR we assume the same detector acceptance for $\nu_e$ and $\nu_\mu$ neutrinos though detection techniques, backgrounds and systematic errors are very different. With this assumption limits scale as $\sim 1/(N\nu_\mu)^{1/2}$, which means that the projected limits with the DSFMR will be higher by a factor of $(400)^{1/2} \approx 20$ with respect to those with ultimate Neutrino Factory. We use the sensitivity projections in references [6, 30] to scale down the sensitivity reach with DSFMR. The far-away detectors used in reference [6] are of 25 kt fiducial mass. We note that with a 100 kt fiducial mass for the DSFMR detectors the sensitivity reach would be a factor of 10 lower than the one projected in [6].

The sensitivity limits for $\sin^2 2\theta_{13}$ projected for the MINOS [31] and NOνA [5] experiments are also shown in figure 19. Until the very long baseline experiments are built NovA experiment would receive neutrino beam form two targets (each 4 MW) operating interchangeably. For the DSFMR and the Neutrino Factories the running times of 4 years with each, neutrino and antineutrino beams were assumed. The running times with NOνA experiment is assumed 3 years with each, the neutrino and the antineutrino beams, and for MINOS 4 years of running is assumed. The sensitivity projections in figure 19 show that the DSFMR based experiments exceed by far projected sensitivity reach with NOνA experiment (and both MINOS and NOνA experiments are degenerate by the CP violation and the sign of $\Delta m^2$ parameter). The sensitivity limits with DSFMR are much below that of with Neutrino Factory but the DSFMR can be put into the operation many years ahead and improving greatly the current neutrino experiments at Fermilab. We must stress, however, that Neutrino Factory has great advantage over the DSFMR as it also allows study the $\nu_e \rightarrow \nu_\mu$ oscillations. Consequently, Neutrino Factory should certainly be considered as the successor to the DSFMR.

The detector choice depends on the neutrino energy, which in turn depends on the energy of the proton beam. In figure 20 we show the mean neutrino beam energy as a function of the proton beam energy. The higher the energy of the neutrino the denser the detector can be used. This is very important because as pointed out in [32] it allows for the neutrinos in the 20 GeV range use the iron based calorimeters, saving space while increasing the fiducial mass. Most of the current neutrino experiments apply low-density medium such as water or the liquid scintillator. Such an approach requires large detector volumes for a fiducial mass necessary to satisfy the detection efficiency. The iron based calorimeters are much simpler to build and they tend to have lower cost and easier operations. The DSFMR based neutrino experiments can only meet the ultimate limit expectations of the Neutrino Factory if the fiducial size of the far-away detectors is considerably increased. Using the iron-based calorimeters certainly facilitates such undertaking.

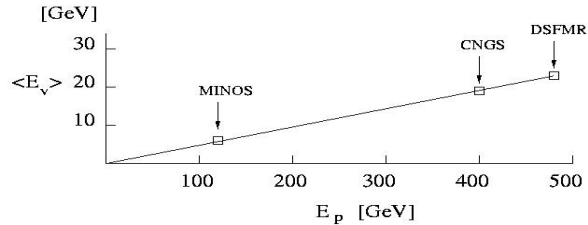

Fig. 20 Mean neutrino energy for MINOS ($E_p = 120$ GeV), CNGS ($E_p = 400$ GeV) and projected for DSFMR ($E_p = 480$ GeV)

## 3. DSMR as pre-injector to VLHC

### 3.1 A layout of scaled-down VLHC in Chicago-land area

A possible location of new VLHC accelerator in the Chicago area is shown in figure 21.

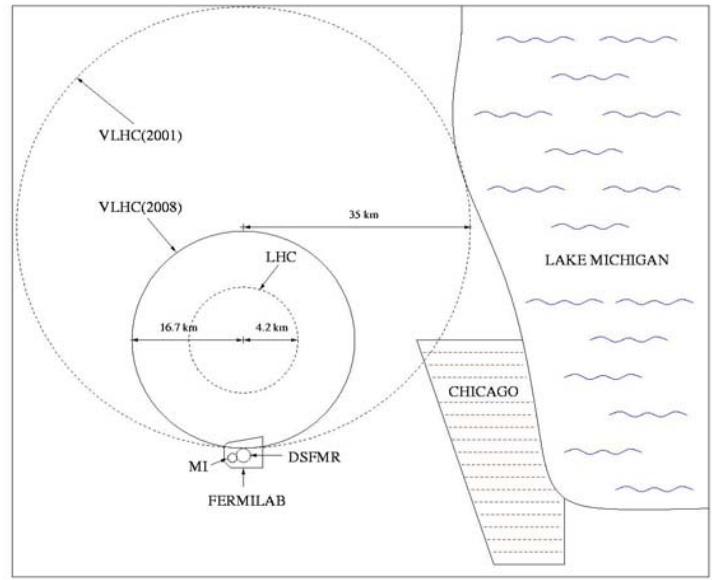

Fig.21 Possible location of VLHC ring in Chicago area. The rings
Of VLHC-2001 and LHC are also shown for comparison.

At the time the VLHC proposal [13] was conceived the adopted guiding principle was building an accelerator in a tunnel of a largest feasible circumference to study the proton-proton collisions at as high as possible energy. This approach may have, however, backfired as it has lead to a project that may have been much too difficult, too expensive, and of too large a scale to manage. Assuming use of the LHC type magnets in the final VLHC accelerator stage the collision energy of 56 TeV (4 times the LHC) is achieved in a circumference of 106 km. The new VLHC ring is far away from geologically difficult areas such as Troy Bedrock Valley, the Sandwich Fault, the Michigan Lake, and it does not interfere with City of Chicago. This makes tunnel construction more feasible from the civil engineering point of view, and more likely acceptable by the populace.

### 3.2 Injection scheme and some basic parameters of VLHC

The arrangement of VLHC rings relative to the DSFMR is shown in figure 22. The VLHC tunnel will host two accelerator rings, the Low Energy Ring (LER) and the High Energy Ring (HER), a concept first developed for the luminosity upgrade at LHC [16]. Two 0.5 TeV proton beams from DSFMR will be simultaneously stacked in the LER ring. Both the LER and the HER rings use two-bore magnets. After stacking is complete the energy of both LER beams will ramp to 7.5 TeV, and then these beams will simultaneously transfer to the HER ring. The two beams in the HER ring will then ramp to the ultimate VLHC energy. At present we assume the HER ring will use the LHC-type 8 Tesla magnets (VLHC I). There is, however, a long-term but realistic possibility of accelerator magnets in the range of 16 Tesla. Such magnets could be used in the future for the upgrade to VLHC II allowing for 115 TeV of the collision energy, or 8 times that of the LHC.

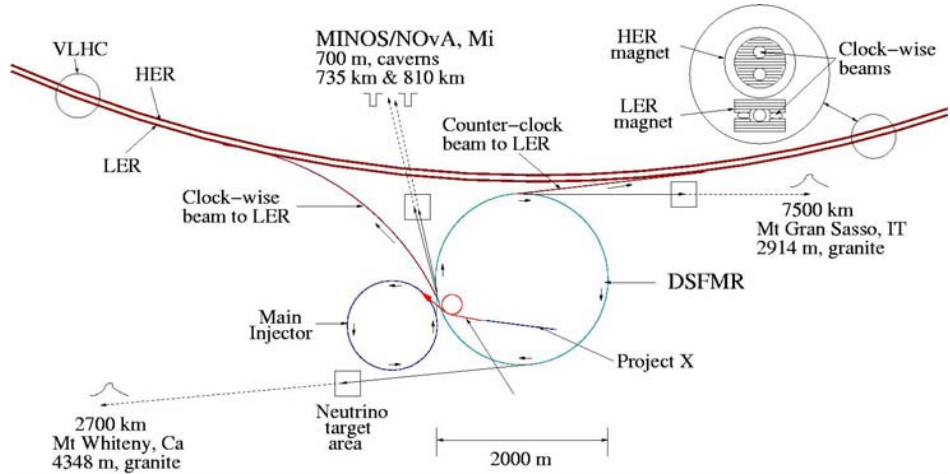

Fig.22 LER and HER VLHC accelerator rings with transfer lines from DSFMR

The proposed arrangement of the new VLHC does not affect the long-baseline neutrino physics program, and it allows its continuation during the construction period as well as after the VLHC was built.

A summary of the basic parameters of all synchrotrons for the Fermilab accelerator complex upgrade is presented in Table 1. We assume that proton beam formation prior to the injection into the Main Injector will use either the present system (0.4 GeV linac with 8 GeV Booster), or a new 8 GeV H$^-$ linac as proposed for the Project X [11].

Table 1 Basic parameters of present and proposed new synchrotrons at Fermilab

| Synchrotron | Circumference [km] | Injection/Extraction Energy [GeV] | Max. B-Field [T] | Ramp rate [T/s] | Cycle Time [s] |
|---|---|---|---|---|---|
| Booster | 0.474 | 8 | 0.7 | 7 | 0.2 |
| Main Injector | 3.4 | 8 / 120 | 1.8 | 2 | 1.4 |
| DSFMR | 6.8 | 48 / 480 | 2.0 | 2 | 2 |
| LER | 106 | 480 / 7200 | 2.0 | 0.01 | 200 |
| VLHC I | 106 | 7200 / 28800 | 8.0 | 0.005 | 1600 |
| VLHC II | 106 | 7200 / 57600 | 16.0 | 0.005 | 3200 |

We addressed the fast-cycling superconducting magnets for DSFMR in Chapter 2.2. The LER accelerator would use magnets based on the VLHC-Low Field design [13], and the HER accelerator in the VLHC I would use the LHC type magnets.

The simultaneous acceleration and then the subsequent simultaneous transfer of the two beams into the next accelerator stage will allow shorten considerably the overall VLHC cycle. This is very important because due to a very high cost of the refrigeration power the beam stacking and acceleration periods when the physics data are not produced should be minimized.

The proton beam intensities at various stages and the projected VLHC luminosity relative to that of LHC are given in Table 2. The beam stacking stops at the LER stage. The efficiency of the beam stacking and transfer between the accelerators was assumed to be at 90% level for each stage. The $n_b$ is the number of bunches, and $N_p$ is the number of protons in a bunch.

Table 2 Projected proton intensity at VLHC as compared to LHC

|  | $n_b$ | $N_p$ / bunch | $N_p$ / ring |
|---|---|---|---|
| Main Injector | 498 | $1.08 \; 10^{11}$ | $5.4 \; 10^{13}$ |
| DSFMR | 996 | $0.97 \; 10^{11}$ | $9.7 \; 10^{13}$ |
| LER | 14940 | $0.87 \; 10^{11}$ | $13 \; 10^{14}$ |
| VLHC | 14940 | $0.78 \; 10^{11}$ | $11.7 \; 10^{14}$ |
| LHC | 2808 | $1.15 \; 10^{11}$ | $2.9 \; 10^{14}$ |

The p-p collider luminosity L is $\sim n_b \times (N_p)^2$. Consequently the projected luminosity of the VLHC relative to that of the LHC is: $(14940/2808) \times (0.78/1.15)^2 = 2.4$. This is achieved with current Fermilab pre-injection system to the Main Injector. So, the VLHC I in addition to 4 times higher collision energy will also have 2.4 times higher luminosity as compared to LHC. If the Project X was implemented, however, the number of protons in the Main Injector would rise to $1.4 \; 10^{14}$, and so the VLHC luminosity would be $\sim 6$ times higher then that of the LHC.

## 4. Summary and conclusions

At present any new truly large-scale HEP project must wait until physics data coming from the LHC get sorted out. The LHC is well set to investigate the Higgs up to 0.8 TeV mass, well beyond the expectations of the Standard Model. The determination of the Higgs mass may be the key to the prospect for the CLIC/ILC as well as for the Muon Collider. If the mass Higgs turns out to be only moderately high the LHC will be able to examine it rather thoroughly. On the other hand if the Higgs mass turns out to be very high, or not even observed at LHC, the required collision energy for the CLIC/ILC as well as that for the Muon Collider may be beyond their technological feasibilities contemplated at present.

It is of utmost importance to continue the experimental high-energy particle physics program in the US during the LHC era which will be characterized for some time by uncertainty about options for the future of HEP. From all the US laboratories it is the Fermilab that has a unique opportunity to embark on a research program that is both very important to the high-energy particle physics and also truly complementary to that of the LHC. The search for the neutrino oscillations in "7500 km + 3000 km" baselines with DSFMR can be certainly viewed as such a program. The achievable neutrino theory parameters with DSFMR will certainly further advance understanding of the neutrino physics and provide a solid background for the future Neutrino Factory. In addition, if a new physics will require collision energy much higher than that of the LHC the DSFMR will constitute a first block in constructing the VLHC.

The DSFMR project does not require to carry-out the R&D effort on a large scale. The magnet R&D and the prototyping is straightforward, inexpensive, and it can be accomplished in a time span of 4-5 years. Because of necessity to implement as soon as possible strong high-energy physics program in the US (after the Tevatron closing) the new project should demonstrate its ability to be successfully built within next 8-10 years, and to be of a moderate cost in the same time. We believe that the DSFMR project can be proven to be just that.
`